\documentclass[showpacs,aps,prb,twocolumn,superscriptaddress,10pt]{revtex4-2}
\usepackage{graphicx} 
\usepackage{bm}
\usepackage{color}
\usepackage{xcolor}
\usepackage{amsmath}
\usepackage{amssymb}
\usepackage{enumerate}
\usepackage{xspace}
\usepackage{mathrsfs}
\usepackage{mathptmx}
\usepackage[colorlinks=true,linkcolor=blue,citecolor=blue,urlcolor=blue]{hyperref}
\usepackage{orcidlink}
\usepackage[utf8]{inputenc}

\begin{document}

\title{Anisotropy of the anomalous Hall effect in the altermagnet candidate Mn$_5$Si$_3$ films}
\author{Miina~Leiviskä}
\affiliation{Univ. Grenoble Alpes, CNRS, CEA, Grenoble INP, IRIG-SPINTEC, F-38000 Grenoble}
\author{Javier~Rial}
\affiliation{Univ. Grenoble Alpes, CNRS, CEA, Grenoble INP, IRIG-SPINTEC, F-38000 Grenoble}
\author{Antonín~Bad'ura}
\affiliation{Department of Chemical Physics and Optics, Faculty of Mathematics and Physics, Charles University, Prague, Czechia}
\affiliation{Institute of Physics, Czech Academy of Sciences, Prague, Czechia}
\author{Rafael~Lopes~Seeger}
\affiliation{Univ. Grenoble Alpes, CNRS, CEA, Grenoble INP, IRIG-SPINTEC, F-38000 Grenoble}
\author{Ismaïla~Kounta}
\affiliation{Aix-Marseille University, CNRS, CINaM, Marseille, France}
\author{Sebastian~Beckert}
\affiliation{Institute of Solid State and Materials Physics, TU Dresden, Dresden, Germany}
\author{Dominik~Kriegner}
\affiliation{Institute of Solid State and Materials Physics, TU Dresden, Dresden, Germany}
\affiliation{Institute of Physics, Czech Academy of Sciences, Prague, Czechia}
\author{Isabelle~Joumard}
\affiliation{Univ. Grenoble Alpes, CNRS, CEA, Grenoble INP, IRIG-SPINTEC, F-38000 Grenoble}
\author{Eva~Schmoranzerová}
\affiliation{Department of Chemical Physics and Optics, Faculty of Mathematics and Physics, Charles University, Prague, Czechia}
\author{Jairo~Sinova}
\affiliation{Institute for Physics, Johannes Gutenberg University Mainz, Mainz, Germany}
\author{Olena~Gomonay}
\affiliation{Institute for Physics, Johannes Gutenberg University Mainz, Mainz, Germany}
\author{Andy~Thomas}
\affiliation{Institute of Solid State and Materials Physics, TU Dresden, Dresden, Germany}
\affiliation{Leibniz Institute of Solid State and Materials Science (IFW Dresden), Helmholtzstr. 20, 01069 Dresden, Germany}
\author{Sebastian T. B. Goennenwein}
\affiliation{Department of Physics, University of Konstanz, Konstanz, Germany}
\author{Helena~Reichlová}
\affiliation{Institute of Physics, Czech Academy of Sciences, Prague, Czechia}
\affiliation{Institute of Solid State and Materials Physics, TU Dresden, Dresden, Germany}
\author{Libor Šmejkal}
\affiliation{Institute for Physics, Johannes Gutenberg University Mainz, Mainz, Germany}
\author{Lisa~Michez}
\affiliation{Aix-Marseille University, CNRS, CINaM, Marseille, France}
\author{Tomáš~Jungwirth}
\affiliation{Institute of Physics, Czech Academy of Sciences, Prague, Czechia}
\author{Vincent~Baltz}
\affiliation{Univ. Grenoble Alpes, CNRS, CEA, Grenoble INP, IRIG-SPINTEC, F-38000 Grenoble}
\begin{abstract}
Altermagnets are compensated magnets belonging to spin symmetry groups that allow alternating spin polarizations both in the coordinate space of the crystal and in the momentum space of the electronic structure. In these materials the anisotropic local crystal environment of the different sublattices lowers the symmetry of the system so that the opposite-spin sublattices are connected only by rotations. This results in an unconventional spin-polarized band structure in the momentum space. This low symmetry of the crystal structure is expected to be reflected in the anisotropy of the anomalous Hall effect. In this work, we study the anisotropy of the anomalous Hall effect in epitaxial thin films of Mn$_5$Si$_3$, an altermagnetic candidate material. We first demonstrate a change in the relative N\'eel vector orientation when rotating the external field orientation through systematic changes in both the anomalous Hall effect and the anisotropic longitudinal magnetoresistance. We then show that the anomalous Hall effect in this material is anisotropic with the N\'eel vector orientation relative to the crystal structure and that this anisotropy requires high crystal quality and unlikely stems from the magnetocrystalline anisotropy. Our results thus provide further systematic support to the case for considering epitaxial thin films of Mn$_5$Si$_3$ as an altermagnetic candidate material.
\end{abstract}

\maketitle
\section{Introduction}
The physical properties of magnetic materials are governed by the intrinsic symmetries of their crystal and spin structures, which are represented by symmetry groups. A good example in the realm of magnetotransport properties is the anomalous Hall effect (AHE) \cite{Nagaosa2010}, where a longitudinal electric field $\mathbf{E}$ generates a transversal Hall current $\mathbf{j}_H$ that is perpendicular to the Hall vector ($\mathbf{h}=(\sigma_{zy},\sigma_{xz},\sigma_{yx})^T$, where $\sigma_{ij}$ are the off-diagonal terms of the conductivity matrix) i.e. $\mathbf{j}_H=\mathbf{h}\times \mathbf{E}$. As $\mathbf{h}$ is an axial vector and odd upon time-reversal ($\mathcal{T}$), non-vanishing AHE is only observed in systems where the symmetries allow for such a vector. In ferromagnetic materials the net magnetization is also a $\mathcal{T}$-odd axial vector and thus $\mathbf{h}$ is allowed \cite{Nagaosa2010} while in magnetic materials with compensated order, $\mathbf{h}$ is allowed upon sufficient lowering of the crystal symmetries \cite{Kubler2014,Chen2014,Nakatsuji2015,Nayak2016,Smejkal2020,Smejkal2022aha}. In antiferromagnetic materials \cite{Jungwirth2016, Baltz2018} where the opposite-spin sublattices are connected by translation or inversion, $\mathbf{h}$ is forbidden by symmetry, while in altermagnetic materials \cite{Hayami2019, Yuan2020, Smejkal2022alt} this connecting symmetry is rotation, which can allow for $\mathbf{h}$ although spin orientation along specific crystal axes can still forbid $\mathbf{h}$ by symmetry \cite{Feng2022, Tschirner2023, GonzalezBetancourt2023, Kluczyk2023, Reichlova2020, Kounta2023, Han2024, Badura2023}. We note that to observe AHE, a weak non-zero spin-orbit coupling is always required to transmit lattice information to the conduction electrons even though the origin of the high signal amplitude are the intrinsic symmetries and thus non-relativistic. The lowered symmetry in altermagnets stems from the two sublattices being inequivalent in terms of the local crystal environment. Altermagnets are characterised by alternating spin-splitting of the energy bands \cite{Hayami2019, Yuan2020, Mazin2021, Smejkal2022alt, Fedchenko2023, Lee2023, Osumi2023, Reimers2023, Sattigeri2023} and they show various core spintronic properties such as the AHE presented above, the Nernst effect (thermal counterpart of the AHE) \cite{Zhou2023, Badura2024, Han2024nernst}, spin-polarized current generation \cite{Naka2019, Ma2021, Bose2022, Bai2022, Karube2022}, as well as giant and tunnel magnetoresistances \cite{Chi2023, Smejkal2022gmr, Xu2023} in heterostructures.

In polycrystalline soft ferromagnets, $\mathbf{h}$ is allowed by symmetry for any magnetization orientation and therefore the AHE is proportional to the component of the magnetization that is parallel to $\mathbf{h}$, i.e.  to $M_s$cos$\theta$, where $M_S$ is the saturation magnetization and $\theta$ is the angle between the external field and the film normal, assuming that the field is strong enough to saturate the magnetization. Exceptions to this behavior can arise in crystalline ferromagnets \cite{Limmer2006, Roman2009} or ferromagnets with strong magnetocrystalline anisotropy, which causes a lag between the external field and the magnetization. In altermagnets, specific N\'eel vector ($\mathbf{L}$) orientations can either allow or exclude the Hall vector $\mathbf{h(\mathbf{L})}$ by symmetry \cite{Smejkal2022aha, Feng2022, GonzalezBetancourt2023} so that anisotropic AHE beyond $M_s$cos$\theta$ can arise.

Anisotropic AHE beyond $M_s$cos$\theta$ in compensated magnets constitutes a hallmark of altermagnetism, as has been recently demonstrated in RuO$_2$ \cite{Feng2022} (intrinsic d-wave altermagnet, i.e. the electronic structure displays four lobes with alternating spin-polarization) and MnTe showing spontaneous AHE \cite{GonzalezBetancourt2023} (intrinsic g-wave, which acquires a d-wave character in the presence of spin-orbit coupling). Both RuO$_2$ and MnTe owe their altermagnetism to the presence of non-magnetic atoms that render the local crystal environments of the sublattices anisotropic \cite{Smejkal2022alt}. The saturation of $\mathbf{L}$ in RuO$_2$ and MnTe requires high fields of 68 \cite{Feng2022, Tschirner2023} and 4 T \cite{GonzalezBetancourt2023}, respectively, so for further studies on the anisotropy of AHE in altermagnets a new material candidate where $\mathbf{L}$-saturation is realized at smaller fields would be ideal. 
\begin{figure}[ht!]
\includegraphics[width=0.49\textwidth, trim=0 0 0 0]{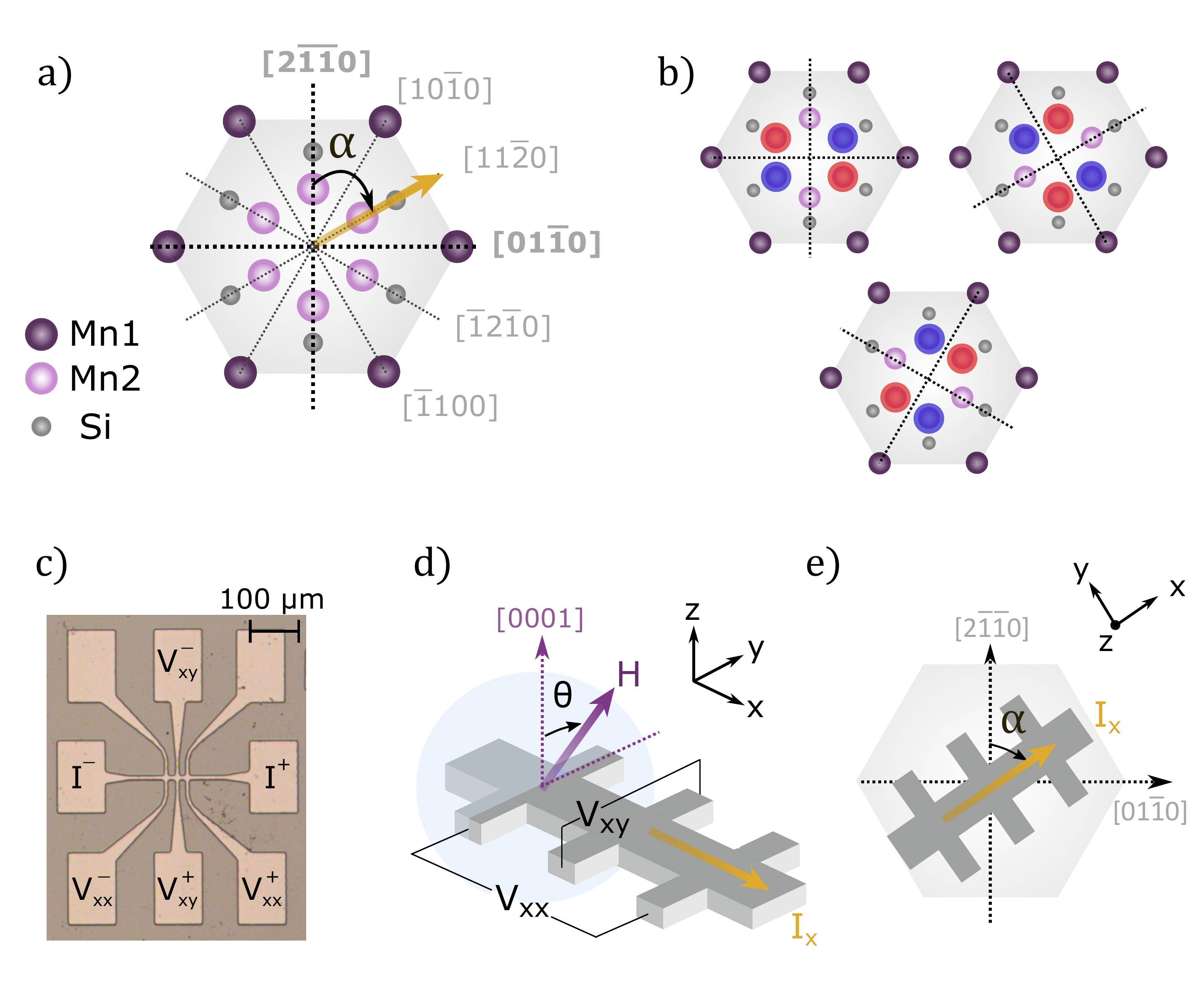}
\caption{(Color Online) (a) The hexagonal unit cell of epitaxial thin films of Mn$_5$Si$_3$. (b) The three options of distributing the two magnetically ordered opposite-spin sublattices of Mn2-atoms (highlighted in red and blue). These three possible quadrupole domains yield six possible magnetic domains when accounting for the reversal of the N\'eel vector.
(c) Optical image of a typical Hall bar device. (d) Illustration of a Hall bar with the angle $\theta$ defining the external field $H$ orientation relative to the film normal z, which coincides with the [0001] crystal axis of Mn$_5$Si$_3$. (e) Top view of a Hall bar showing the angle $\alpha$, which refers to the orientation of the current channel $I_{x}$ relative to the crystal axis $[2\bar{11}0]$. Varying $\alpha$ is achieved through the fabrication of a set of Hall bars oriented along several crystal directions.}
\label{fig:1}
\end{figure}

Recently, high crystal quality  thin films of Mn$_5$Si$_3$ have emerged as such candidate \cite{Reichlova2020, Kounta2023, Han2024, Badura2024, Han2024nernst} as the collinear compensated phase of Mn$_5$Si$_3$ with a hexagonal unit cell and four magnetic Mn atoms on the Mn2 sites in a checkerboard arrangement of the sublattices (shown in Figure \ref{fig:1}(a)) fulfills the symmetry criteria for altermagnetism \cite{Reichlova2020}. The altermagnetic phase has been supported by the experimental observations of both anomalous Hall \cite{Reichlova2020, Han2024} and Nernst effects {\cite{Badura2024, Han2024nernst}}. Note, that in the bulk phase of Mn$_5$Si$_3$ the hexagonal unit cell is not preserved during the magnetic ordering and the opposite-spin sublattices become connected by translation, making bulk Mn$_5$Si$_3$ antiferromagnetic \cite{Lander1967,Menshikov1990,Brown1992,Brown1995}. However, when grown as epitaxial thin films \cite{Kounta2023} the substrate stabilizes the hexagonal unit cell of Mn$_5$Si$_3$ over the entire temperature range \cite{Reichlova2020}, fulfilling the symmetry requirements for making epitaxial Mn$_5$Si$_3$ a potential candidate for the altermagnetic phase with a d-wave symmetry. In good accordance with the theoretical predictions and the structural properties, epitaxial Mn$_5$Si$_3$ exhibits a sizable spontaneous AHE despite a vanishingly small net magnetization \cite{Reichlova2020}. We note that vanishing net magnetization and non-vanishing AHE are also observed in non-collinear antiferromagnets with broken time reversal symmetry, such as Mn\textsubscript{3}Sn and Mn\textsubscript{3}Ge \cite{Chen2021}. However, previous studies \cite{Reichlova2020, Kounta2023, Han2024, Badura2024, Han2024nernst} through theoretical and experimental arguments show that non-collinear order in epitaxial Mn\textsubscript{5}Si\textsubscript{3} films is unlikely.

In this work we investigate the anisotropy of AHE in epitaxial Mn$_5$Si$_3$ in terms of the external field orientation relative to the crystal structure. We show that a systematic rotation of the field in different crystal planes inflicts systematic changes on the magnetotransport properties, indicating that we are able to control the relative $\mathbf{L}$-orientation with experimentally reasonable external field strengths of the order of 2 T. Moreover, we show that changing the relative $\mathbf{L}$-orientation yields anisotropic behavior of the AHE, beyond cos$\theta$, that requires a high crystal quality of the epitaxial films and is unlikely related to the magnetocrystalline anisotropy, which further demonstrates that the properties of epitaxial Mn$_5$Si$_3$ closely match those expected for altermagnetic materials. This is in line with previous findings in Mn$_5$Si$_3$ that predict collinear compensated hexagonal phase and demonstrate strong spontaneous anomalous Hall \cite{Reichlova2020, Han2024} and Nernst {\cite{Badura2024, Han2024nernst} effects with a clear influence of crystallinity \cite{Kounta2023} and Mn-content \cite{Han2024nernst}.

\begin{figure*}[tb]
\includegraphics[width=0.99\textwidth]{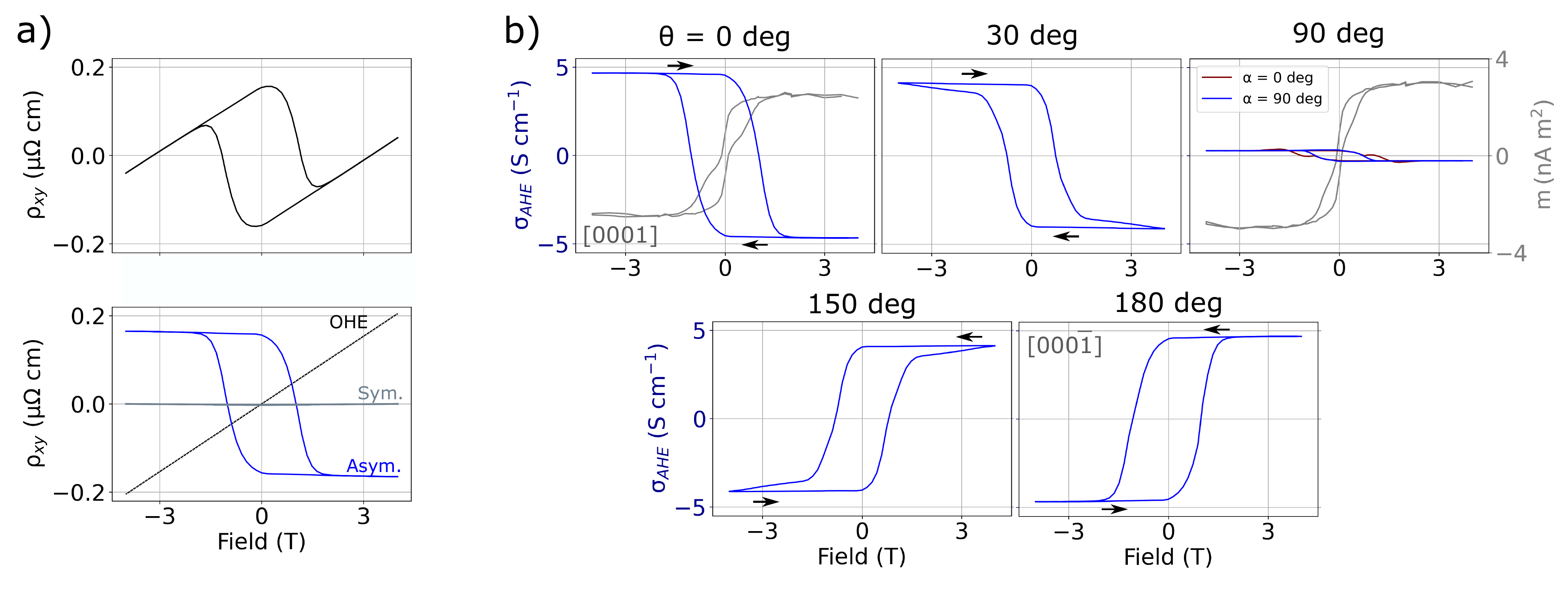}
\caption{(Color Online) (a) Disentangling the different contributions (symmetric-in-field $\rho_{xy, sym}$, antisymmetric-in-field $\rho_{xy, asym}$, and ordinary Hall effect $\rho_{OHE}$) to the field-dependence of transverse resistivity $\rho_{xy}$. The signal of interest here is $\rho_{xy, asym}$, relating to the anomalous Hall effect (AHE) resistivity $\rho_{AHE}$. (b) The hysteresis loops of the corresponding AHE conductivity $\sigma_{AHE}={\rho_{AHE}}/{\rho_{xx}^2}$ at selected external field orientations $\theta$ at 110 K. The sweep direction is indicated by the small arrows. The $\sigma_{AHE}$ is at maximum at $\theta=$ 0 deg (field out-of-plane) and decreases when the field rotates away from it, almost vanishing at $\theta=$ 90 deg (field in-plane). This is in contrast with the magnetic moment $m$ and corresponding weak magnetization $M$, which is isotropic for $\theta=0$ and 90 deg. For $\theta=$ 90 deg,  the saturation fields are similar and below 2 T for both $\alpha = 0$ and 90 deg, pointing to similar magnetocrystalline anisotropy. Note that we have removed the linear diamagnetic background in the hysteresis loops of $m$.} 
\label{fig:2}
\end{figure*}

\section{Methods}
The epitaxial thin films of Mn$_5$Si$_3$ are grown on Si(111) substrates (thickness $\sim$ 250 $\mu$m,  $\rho>$ 10 k$\Omega$ cm) by molecular beam epitaxy using optimized growth parameters for realizing films with high crystalline quality and maximum Mn$_5$Si$_3$ content \cite{Kounta2023}. A few nm-thick MnSi phase acts as a thin seed layer aiding the nucleation of Mn$_5$Si$_3$ on Si(111) \cite{Kounta2023}. Unless otherwise specified, the films used in this work contain 96 \% Mn$_5$Si$_3$ and 4 \% MnSi and they have a thickness of $\sim$17 nm. The high crystallinity and the crystal orientation of the films have been verified using transmission electron microscopy (TEM) in our previous works \cite{Reichlova2020, Kounta2023}. The epitaxial relationship between the Si substrate, MnSi seed layer and Mn$_5$Si$_3$ film is Si(111)$[1\bar{1}0]$//MnSi(111)$[\bar{2}11]$//Mn$_5$Si$_3$(0001)$[01\bar{1}0]$. The films were patterned into microscopic Hall bars shown in Figure \ref{fig:1}(c) using optical UV lithography and subsequent ion beam etching. The orientations of the Hall bar current channels relative to the crystal axis $[2\bar{1}\bar{1}0]$ are defined by the angle $\alpha$ as illustrated in Figure \ref{fig:1}(e). We measure simultaneously the longitudinal ($V_{xx}$) and transversal ($V_{xy}$) voltages (denoted in Figure \ref{fig:1}(c,d)) as a function of the external field strength and orientation to obtain the corresponding field- and angular-dependences of the longitudinal ($\rho_{xx}$) and transversal ($\rho_{xy}$) resistivities. We apply the external field always in the (yz) plane, i.e. the plane perpendicular to the current channel. Its exact orientation relative to the [0001] crystal axis (film normal) is given by the angle $\theta$, as illustrated in Figure \ref{fig:1}(d). Note that we will demonstrate later on in the text that the important parameter causing the anisotropic AHE is the crystal plane in which the external field rotates rather than the exact crystal orientation of the current channel - we alter the Hall bar orientation simply to ensure that the external field and the current channel are always perpendicular to each other unless otherwise stated. The current density is $3\times10^{5}$ A cm$^{-2}$ throughout this work. For all the measurements presented in this work, the temperature was kept at 110 K, which is between the two critical temperatures of our films  \cite{Reichlova2020,Kounta2023}. At this temperature the films exhibit characteristic features of an altermagnetic phase, i.e. a sizable spontaneous AHE despite a vanishingly small net magnetization \cite{Reichlova2020}.

\section{Results and discussions}
First, we study how the AHE resistivity ($\rho_{AHE}$) evolves when the external field orientation, parameterized by $\theta$, is changed. We consider a Hall bar where the external field rotates in the (0$\bar{1}$10) plane ($\alpha$ = 90 deg). We have first measured the hysteresis loops of $\rho_{xy}$ for selected $\theta$, from which a linear $\rho_{OHE}$ background (slope $\sim$ 0.045 $\mu\Omega$ cm $T^{-1}$), originating from the ordinary Hall effect has been removed and only the antisymmetric-in-field part of the signal is used: $\rho_{AHE}=\rho_{xy, asym}=(\rho_{xy}(H)-\rho_{xy}(-H))/2$. The symmetric part $\rho_{xy, sym}=(\rho_{xy}(H)+\rho_{xy}(-H))/2$, is negligible in our sample at the specific measurement temperature supporting the collinear spin configuration \cite{Badura2023}. The different contributions to $\rho_{xy}$ are summarized in Figure \ref{fig:2}(a). For the remaining of the paper, we will convert $\rho_{AHE}$ to AHE conductivity $\sigma_{AHE}={\rho_{AHE}}/{\rho_{xx}^2}$, which is used to characterize the intrinsic effects ($\rho_{xx}$ at 110 K is $\sim$ 154 $\mu\Omega$ cm).

The hysteresis loops of $\sigma_{AHE}$ for various $\theta$ are shown in Figure \ref{fig:2}(b). We observe that the $\sigma_{AHE}$ at saturation ($\sigma_{AHE, max}$) is maximum when the field is close to the out-of-plane [0001] axis ($\theta=0$ deg) and vanishes when the field is close to the in-plane [2$\bar{1}\bar{1}0$] axis ($\theta=90$ deg). As expected from the previous AHE study on the epitaxial Mn$_5$Si$_3$ \cite{Reichlova2020}, 180 degree rotation of this field changes the polarity of $\sigma_{AHE}$. The evolution of $\sigma_{AHE}$ for the field orientations between [0001] and [000$\bar{1}$] is better visible in Figure \ref{fig:3} where we directly show $\sigma_{AHE}$ as a function of $\theta$ for a 4 T external field. Between [0001] and [000$\bar{1}$] $\sigma_{AHE}$ shows step-like behavior: when the field is oriented close to either [0001] or [2$\bar{1}\bar{1}0$] (in-plane) crystal axes the slope is considerably more flat than in-between these directions. We note that the backward sweep does not trace the forward sweep because the saturation is not reached at 4 T at all field orientations as seen in Figure \ref{fig:2}(b). 
\begin{figure}[hb!]
\includegraphics[width=0.45\textwidth]{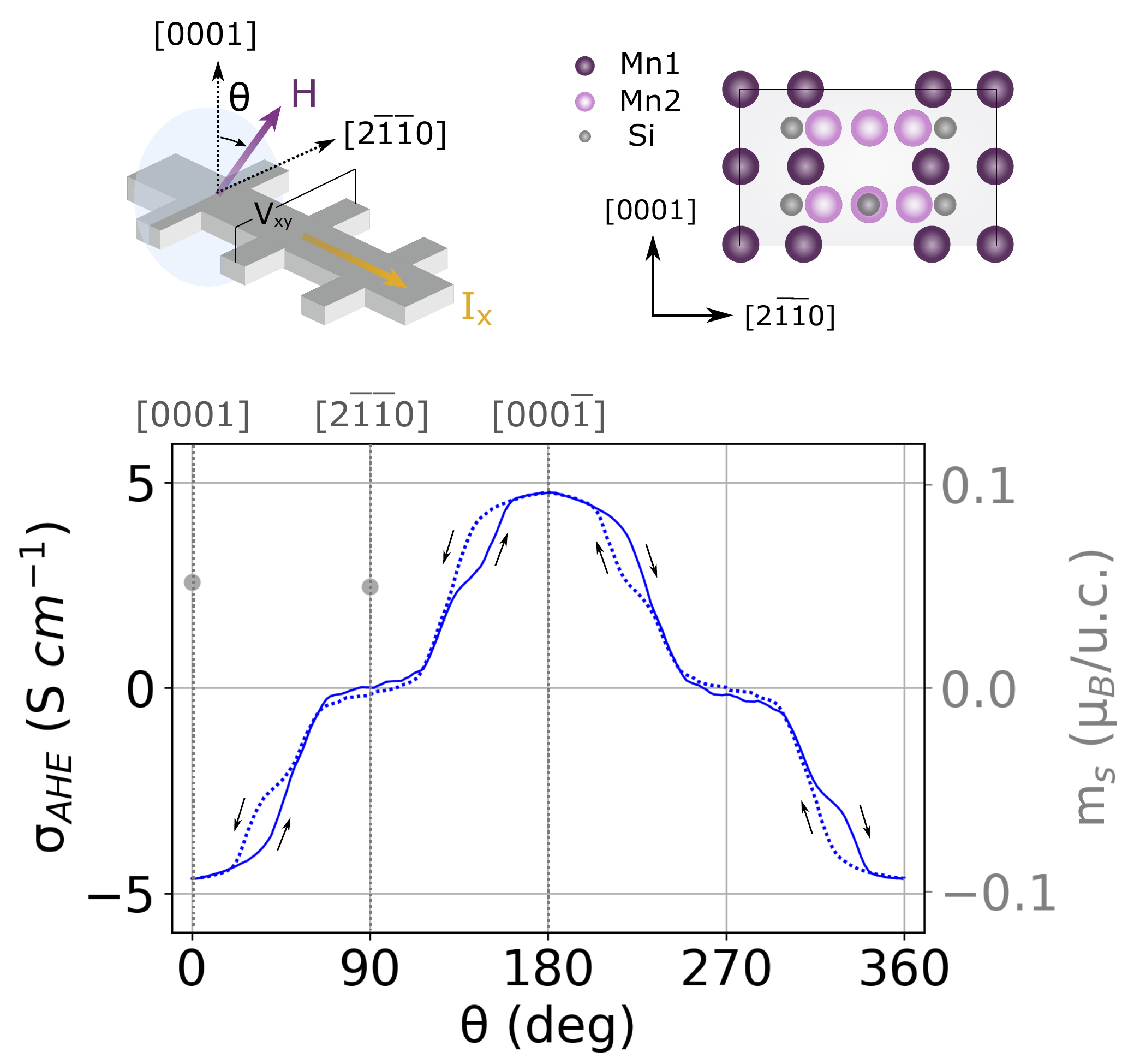}
\caption{(Color Online) The AHE conductivity $\sigma_{AHE}$ as a function of the orientation $\theta$ of a 4 T external field, at 110 K (full line), for a current applied along the $[01\bar{1}0]$  direction ($\alpha$ = 90 deg). The forward sweep is indicated by the full line and the backward sweep by the dotted line. The AHE signal does not correlate with the $\theta$-dependence of the weak saturation magnetization $M_s$ of the hysteretic component (circles), which is isotropic at $\theta=0$ and 90 deg. Inset: measurement geometry recalled and crystal structure of the (01$\bar{1}$0) plane in which the external field $H$ is rotated.}
\label{fig:3}
\end{figure}
\begin{figure}[ht!]
\includegraphics[width=0.49\textwidth, trim=0 0 0 0]{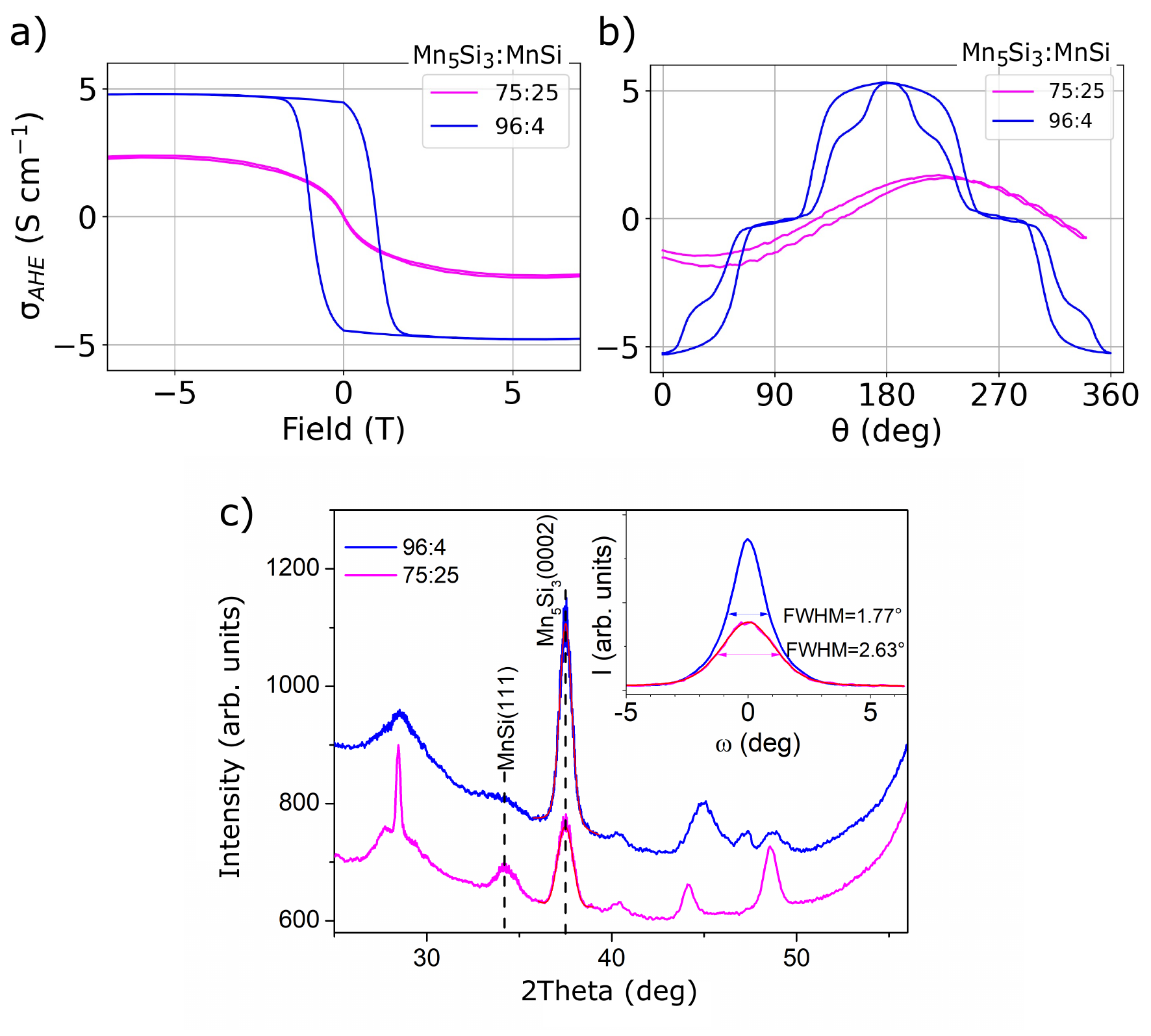}
\caption{(Color Online) (a) The field strength and (b) $\theta$-orientation dependence (at 2 T) of $\sigma_{AHE}$ at 110 K for epitaxial films with 96:4 and 75:25 Mn$_5$Si$_3$:MnSi ratios. The former is an altermagnetic candidate and shows sizable AHE with large coercivity and unconventional step-like $\theta$-dependence. The latter is expected not altermagnetic \cite{Kounta2023} and it shows smaller AHE with vanishing coercivity and conventional isotropic $\theta$-dependence. (c) X-ray diffraction (XRD) profiles generated from the integrated intensities for equal radial distance of the 2D-XRD patterns of films with two different Mn$_5$Si$_3$:MnSi ratios. Inset: rocking curves measured on the Mn$_5$Si$_3$-(0002) reflections.}
\label{fig:4}
\end{figure}
\begin{figure}[ht!]
\includegraphics[width=0.49\textwidth, trim=0 0 0 0]{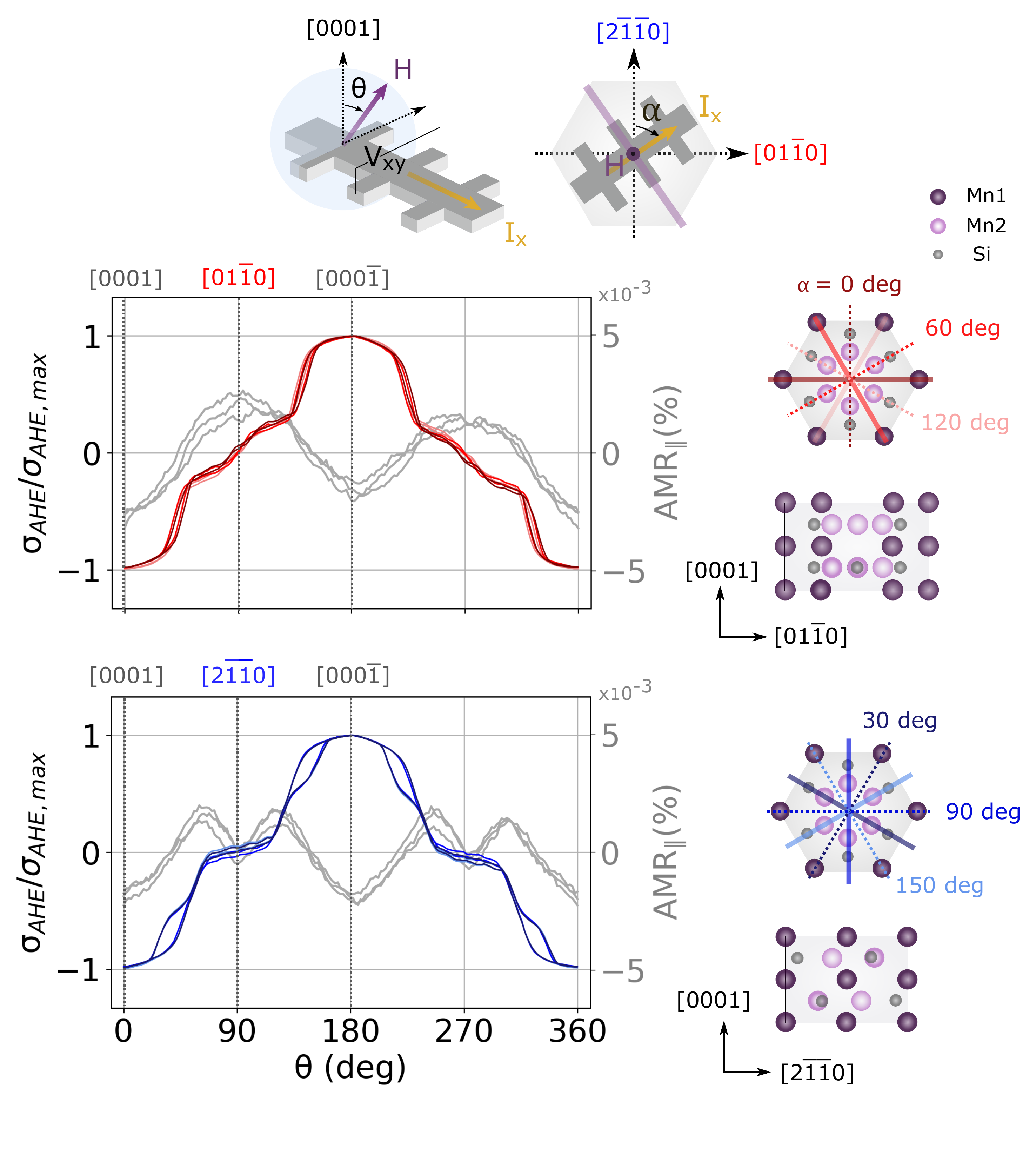}
\caption{(Color Online) The normalized AHE conductivity $\sigma_{AHE}/\sigma_{AHE,max}$ and the longitudinal anisotropic magnetoresistance (AMR$_{\parallel}$) when the external field is rotated in different crystal planes. In both cases, two distinct behaviors can be observed depending on whether the crystal plane is equivalent to (0$\bar{1}$10) or (2$\bar{1}\bar{1}$0). Here, the field strength is 4 T and the temperature is 110 K. The inserts recall the geometry used and the planes in which the field is rotated.}
\label{fig:5}
\end{figure}

To unravel the possible origin of this anisotropic behavior of the AHE, we will first study the option of weak magnetization giving a contribution. Magnetic characterization was carried out using a superconducting quantum interference device (SQUID) magnetometer (Quantum Design MPMS XL 5.0). Before measurement the sample was cleaned with acetone, alcohol, and deionized water, and them mounted (in two different orientations) on a sample holder made of two quartz capillaries with two dots of glue. Analogously to the AHE experiments, we have measured the sample magnetization $M$ along the external field for two field orientations $\theta=0$ and 90 deg as shown in Figures \ref{fig:2}(b) and \ref{fig:3}. After removing a linear diamagnetic background, we observe that the magnetization loops contain two components: a non-hysteretic part that relates to the substrate and the sample holder, and a hysteretic part that we attribute to the Mn$_5$Si$_3$ film as the magnitude of its coercivity agrees with that of the AHE-signal. This hysteretic component suggests a slight spontaneous canting of the magnetic sublattice moments and can play a role in the reversal mechanism of the N\'eel vector. We convert the moment $m$ of the hysteretic component to magnetization using $M \lbrack \mu_B/u.c\rbrack = m \cdot V_{u.c.}/(V_{film}\cdot\mu_B)$ where the volume of the unit cell is $V_{u.c.} = a^2c\sin(60^{\circ})$, where a = 6.93 Å and c = 4.77 Å for our films at 110 K \cite{Reichlova2020}. The volume of the magnetic film is $V_{film} \sim 17$ nm$\times 5 $ mm$\times4$ mm. We observe that the magnetization of the hysteretic component is $\sim0.05$ $\mu_B/$u.c., and this value for $\theta=0$ and 90 deg is similar. While still being vanishingly small this value slightly exceeds the previously reported values \cite{Reichlova2020, Han2024}, which likely stems from the slight variations in the magnetic properties between the different thin films. We note that the non-zero AHE despite the vanishing net magnetization could also arise from non-collinear AFM ordering in our films \cite{Nakatsuji2015, Nayak2016} but as previously shown this arrangement is unlikely in our material at the considered temperature \cite{Reichlova2020}. Alternatively,} the behavior shown in Figure \ref{fig:2}(b) would also be expected for a ferromagnet as $M_z\sim 0$ at $\theta = 90$ deg, while the wide plateau in Figure \ref{fig:3} could be a consequence of the magnetocrystalline anisotropy. We will now discuss further experiments to show that the anisotropic AHE is more likely to originate from the presumed \cite{Reichlova2020, Kounta2023, Han2024, Badura2024, Han2024nernst} altermagnetic phase of our films.

First, we have measured the AHE in Mn$_5$Si$_3$ films with a larger proportion of MnSi (Figure \ref{fig:4}), namely 75 \% Mn$_5$Si$_3$ and 25 \% MnSi, but with a similar thickness of $\sim$16 nm and longitudinal resistivity $\rho_{xx}$ of $\sim$190 $\mu\Omega$ cm at 110 K (for bare MnSi film, $\rho_{xx}$ is $\sim$ 220 $\mu\Omega$ cm at 110 K). Increasing the MnSi content has been shown to deteriorate the crystallinity of the Mn$_5$Si$_3$ film, leading to a more textured film \cite{Kounta2023}. In the extreme polycrystalline case, altermagnetism can be prohibited as the strain necessary for stabilizing the hexagonal unit cell can relax at the grain boundaries, or if the hexagonal unit cell is retained in the individual grains their different crystal orientations would result in overall cancelling out of the altermagnetic effects. In Figure \ref{fig:4}(c) we show that the 75:25 sample is more mosaic than the 96:4 judging from its wider broadening of the full width at half maximum (FWHM) of the rocking curves around the Mn\textsubscript{5}Si\textsubscript{3}-(0002) reflections (shown in the inset of Figure \ref{fig:4}). This increase in mosaicity seems to be sufficient to result in the magnetotransport properties of this film to be different from those of the 96:4 (Mn$_5$Si$_3$:MnSi) films with higher crystal quality: the field dependence of $\sigma_{AHE}$ of the 75:25 control sample has a reduced amplitude and a vanishing coercivity, while the $\theta$-dependence of the $\sigma_{AHE}$ shows roughly a $\cos\theta$ behavior, pointing to a ferromagnetic character. These results are in stark contrast with the behavior of the 96:4 sample and demonstrate that the AHE with anisotropic $\theta$-dependence and non-zero coercivity requires a high crystal quality of the Mn$_5$Si$_3$ epitaxial films, which is consistent with the presumed \cite{Reichlova2020, Kounta2023, Han2024, Badura2024, Han2024nernst} altermagnetic phase being at the origin of the AHE as it requires the system to have a specific symmetry. Moreover, this shows that the spurious MnSi phase with $T_c\sim$ 40 K is also unlikely origin of the AHE signal as the $\sigma_{AHE}$ should in that case be stronger for the 75:25 sample.

\begin{figure}[ht!]
\includegraphics[width=0.4\textwidth, trim=0 0 0 0]{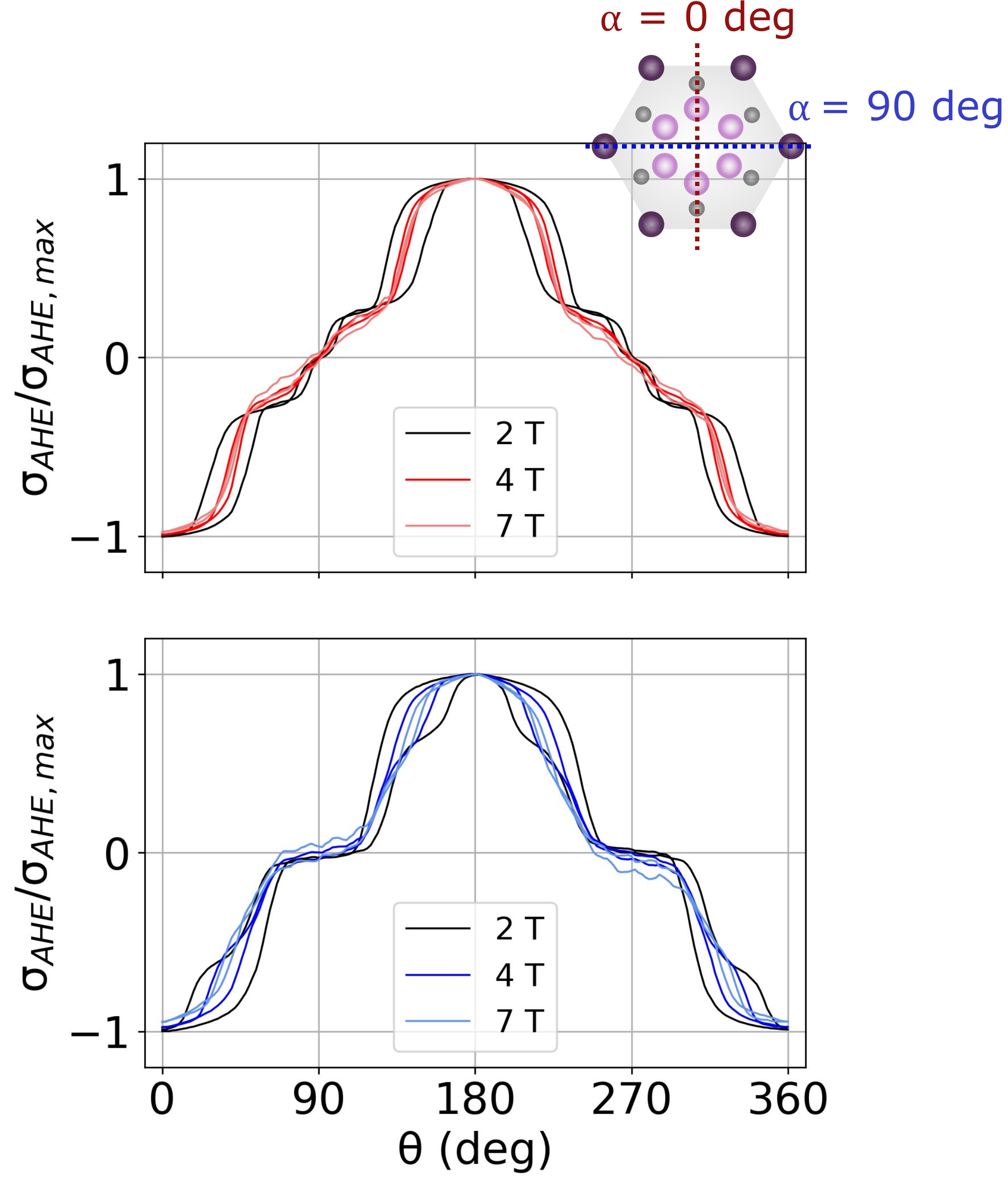}
\caption{(Color Online) The normalized AHE conductivity
$\sigma_{AHE}/\sigma_{AHE,max}$ for two perpendicular Hall bars measured at different field strengths. The inset shows the crystal plane in which the field is rotating while the current channels are along crystal axes perpendicular to these planes. The anisotropy between the two Hall bars persists at 7 T suggesting magnetocrystalline anisotropy as an unlikely origin of the anisotropic behavior.}
\label{fig:7}
\end{figure}

Next, to further study the anisotropy of AHE, we vary systematically the crystal plane in which the external field and in effect the N\'eel  vector $\mathbf{L}$ (confirmed later using the anisotropic magnetoresistance measurements) is rotated and again observe how the $\sigma_{AHE}$ is affected. This was realized by patterning multiple Hall bars with their current channels oriented along different crystal axes. We recall that these orientations are denoted by $\alpha$, which is the angle between the current channel and the crystal axis $[2\bar{1}\bar{1}0]$ as was shown in Figure \ref{fig:1}(e). For each Hall bar we rotate the field in a plane perpendicular to the current channel and thus effectively change the crystal plane in which the field rotates. We have chosen to focus on only the high symmetry crystal axes selecting $\alpha$'s that reflect the hexagonal crystal structure of our material and can be divided into two sets: for $\alpha=0,60,120$ deg the current channel is along [$2\bar{1}\bar{1}0$] or equivalent so the field rotates in the crystal plane ($2\bar{1}\bar{1}0$) or equivalent;
\begin{figure}[ht!]
\includegraphics[width=0.43\textwidth, trim=0 0 0 0]{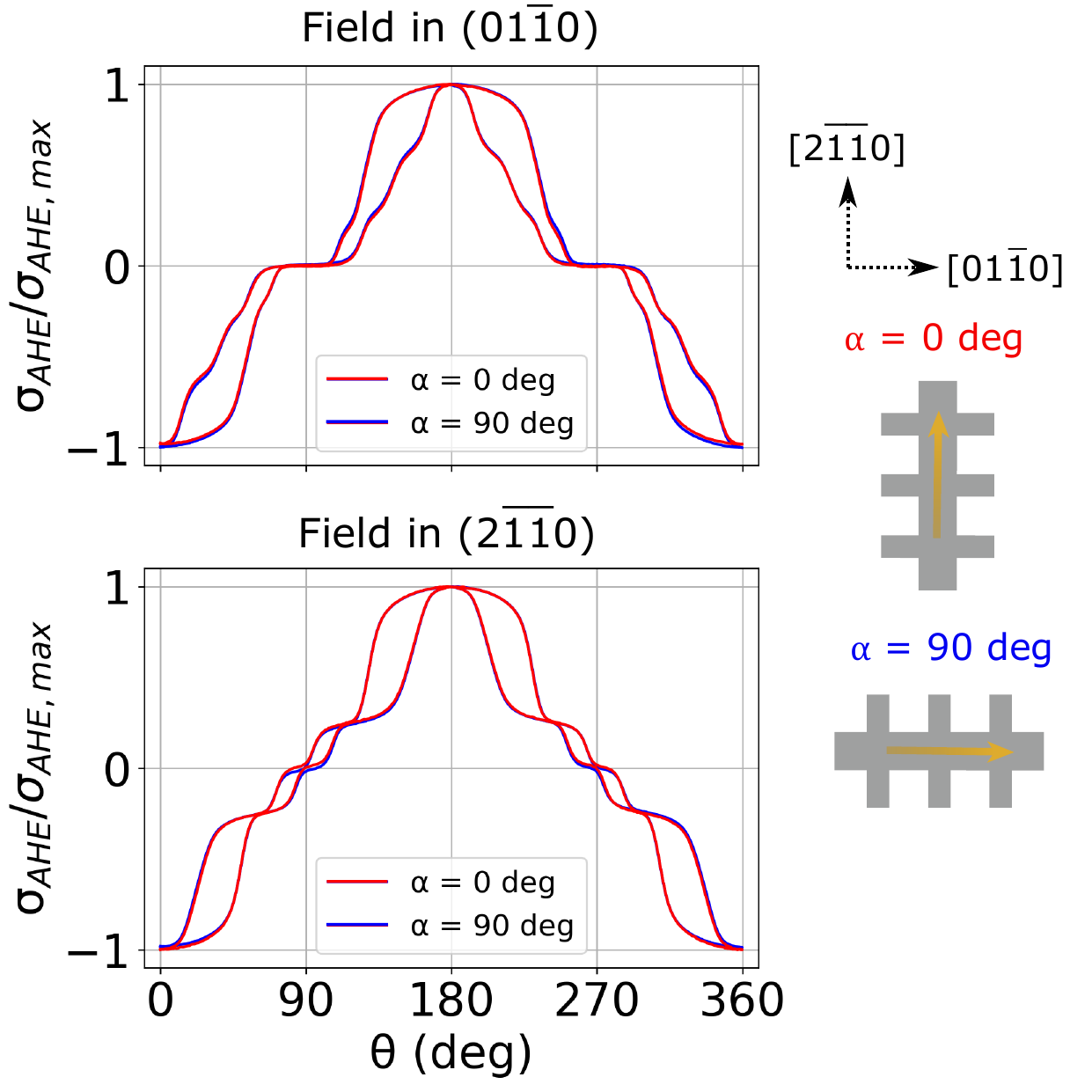}
\caption{(Color Online) Two perpendicular Hall bars measured simultaneously at 110 K with 2 T field. The anisotropic $\theta$-dependence of $\sigma_{AHE}$ relates to the crystal plane in which the field rotates instead of the crystal axis along which the current channel orients.}
\label{fig:6}
\end{figure}
for $\alpha=30,90,150$ deg the current channel is along [$0\bar{1}10$] or equivalent and the field rotates in the crystal plane ($0\bar{1}10$) or equivalent. In Figure \ref{fig:5} we show the $\theta$-dependence of the $\sigma_{AHE}$ (with 4 T field) for the two sets of $\alpha$s. For each set, $\sigma_{AHE}$ is maximum when the field is close to [0001] axis but significant differences arise when the field orients towards the sample plane: for $\alpha= 30, 90, 150$ deg, when the field orients in the [$2\bar{1}\bar{1}0$] direction or equivalent, we observe a wide plateau around $\theta$ = 90 deg, familiar from Figure \ref{fig:3}, while for $\alpha$ = 0, 60, 120 deg when the field orients in the [$0\bar{1}10$] direction or equivalent we observe a steeper linear slope instead. Intermediate angles, like $\alpha= 15, 45,...$ deg show intermediate behavior and provide no further information. We have verified that the anisotropy persists also at 2 and 7 T fields (Figure \ref{fig:7}) . Moreover, the hexagonal symmetry of the AHE signals for the different $\alpha$ suggests that the three possibilities for the quadrupole domains, which N\'eel vector may have a parallel component  (Figure \ref{fig:1}) are likely present in equal populations: if each Mn2 site is on average magnetic, the $\alpha$-dependence of $\sigma_{AHE}$ should reflect the hexagonal symmetry of the crystal structure. If one of these quadrupole domains dominated, lower symmetry for the $\alpha$-dependence of $\sigma_{AHE}$ would be expected. As the net AHE signal does not cancel out implies that the six magnetic domains corresponding to the three quadrupole domains and their reversed N\'eel vector counterparts are not equally populated. We have attempted to modify the population of the quadrupole domains through field-cooling in order to observe a corresponding break in the hexagonal symmetry of AHE signals. We field cooled the sample from 300 to 110 K along two different in-plane directions: [$2\bar{1}\bar{1}0$] and [$0\bar{1}10$] at 1.9 T field, after which we measured the angular $\theta$-dependence at 110 K with a 1.9 T field, for $\alpha=0$ and $90$  deg. We observed the same results as the zero-field cooled case considered in the manuscript  (Figures \ref{fig:5},\ref{fig:6}, and  \ref{fig:7}), suggesting that the field cooling has no influence on the angular dependence of the AHE or the hexagonal anisotropy between the different crystal planes in which the field rotates. 
From this, we infer that the quadrupole domains could not be manipulated with field cooling. In fact, while the magnetic field will influence the Neel order, the response of the non-ordered Mn atoms (Figure \ref{fig:1}) to the field is likely non-trivial and thus altering the population of the different quadrupole domains is also non-trivial. We also note that the presumed quadrupole domains in Mn\textsubscript{5}Si\textsubscript{3} due to some Mn atoms not being ordered (Figure \ref{fig:1}b) differ from other multipole domains, such as the cluster octupole domains in the Mn\textsubscript{3}Ge and Mn\textsubscript{3}Sn non-collinear antiferromagnets \cite{Chen2021, Sugimoto2000}, where all Mn atoms are magnetic and which can be manipulated with an external field.

We have verified that the key factor responsible for the change of behavior between the different $\alpha$s is the crystal plane in which the field rotates rather than the crystal axis along which the current channel is oriented, as observed in Figure \ref{fig:6}. We note that for Figure \ref{fig:6} we have used a different sample of the same material, which explains some of the differences in the $\theta$-dependence of the $\sigma_{AHE}$ compared to the other figures. However, the overall behavior is the same, which confirms the reproducibility of the results discussed here. In summary, the results in Figures \ref{fig:5} and \ref{fig:6} allow us to conclude that the anisotropic behavior of the AHE depends on the $\mathbf{L}$-orientation relative to the hexagonal crystal symmetry of the Mn$_5$Si$_3$ epitaxial films. We also consider magnetocrystalline anisotropy as the origin of the anisotropic AHE unlikely: if we interpret the results shown in Figures \ref{fig:3} and \ref{fig:5} considering that the different crystal planes have strongly different magnetocrystalline energy landscapes, then for $\alpha$ = 30, 90, and 150 deg it is more energetically favorable for $\mathbf{L}$ to remain aligned along a crystal axis yielding $\sigma_{AHE}=0$ at $\theta=90$ deg compared to $\alpha$ = 0, 60, and 120 deg. However, the hysteresis loops measured at $\theta=$ 90 deg for both $\alpha =$ 0 and 90 deg (Figure \ref{fig:2}(b)) superimpose at 4 T, indicating that it is unlikely that at this field $\mathbf{L}$ is lagging behind the external field around $\theta=$ 90 deg for $\alpha$ = 30, 90, and 150 deg more than for $\alpha$ = 0, 60, and 120 deg. Moreover, as shown in Figure \ref{fig:7} we observe the anisotropy in the $\alpha$-dependence of $\sigma_{AHE}$ persisting up to 7 T, which is well above the saturation field of $\sigma_{AHE}$ at any $\theta$, as shown in Figure \ref{fig:2}(b). Along with our demonstration that the anisotropic AHE is crystal quality-dependent and depends on the $\mathbf{L}$-orientation relative to the hexagonal crystal symmetry, these observations provide further support to the main AHE contribution in our Mn$_5$Si$_3$ films being from altermagnetism. These results constitute the main conclusion of our work and they systematically support the previous calculations and experiments on the same system \cite{Reichlova2020}.

Finally, we confirm the ability to manipulate $\mathbf{L}$ with the external field by measuring the anisotropic magnetoresistance (AMR), which is known to depend on the $\mathbf{L}$-orientation. As shown in Figure \ref{fig:5}, we have measured the $\alpha$- and $\theta$-dependence of the longitudinal AMR (AMR$_{||}$), which relates to $\rho_{xx}$ as AMR$_{||}(\theta)={(\rho_{xx}(\theta)-\langle\rho_{xx}\rangle)}/{\langle\rho_{xx}\rangle}$, where $\langle\rho_{xx}\rangle$ indicates $\rho_{xx}$ averaged over all angles. AMR is an even function of the field and known to have two components: a non-crystalline one scaling with the angle between the order parameter and the current direction, and a crystalline one relating to the order parameter orientation relative to the crystal axes \cite{Rushforth2007}. We show the $\theta$-dependence for $\alpha$ = 0, 60, 120 deg and $\alpha$ = 30, 90, 150 deg Hall bars in Figure \ref{fig:5} at conditions identical to the AHE measurements. Again we observe different behaviors for the two sets of $\alpha$, as expected from the crystalline AMR depending on the crystal plane in which the order parameter rotates: for $\alpha$ = 0, 60, 120 deg the dependence follows a $\cos{2\theta}$ behavior while for $\alpha$ = 30, 90, 150 deg higher order even cosines appear in the signal. 

Although the exact N\'eel vector orientation and rotation under the action of the external magnetic field cannot be deduced from the results presented, we wish to open a discussion on the interpretation of the anisotropic AHE in the light of the symmetries of the epitaxial Mn$_5$Si$_3$. This material is considered to be a candidate for the altermagnetic phase \cite{Reichlova2020, Kounta2023, Han2024, Badura2024, Han2024nernst}. First, it should be noted that in our measurement geometry (Figure \ref{fig:1}), the changes in $\sigma_{AHE}$ correspond to changes in the component of $\mathbf{h}$ along the [0001] direction (out-of-plane). In altermagnets, the presence and orientation of $\mathbf{h}$ depend on the $\mathbf{L}$-vector orientation because the latter influences the symmetries of the system. While accurate calculation of the entire $\theta$-dependence of $\mathbf{h(\mathbf{L})}$ is inhibited by the unknown relevant micromagnetic parameters, the value of $\sigma_{AHE}$ for some specific orientations can be inferred from symmetry considerations. More specifically, considering the expected spin structure of the epitaxial Mn$_5$Si$_3$ shown in Figure \ref{fig:1}(a) we can use symmetry analysis to obtain information on the $\mathbf{L}$-vector orientations that give zero $\sigma_{AHE}$ in the xy direction: if $\mathbf{L}$-vector is along [0001] crystal axis, $\mathbf{h}$ is forbidden in any direction due to $C_{2x}$, $C_{2y}$, and $C_{4z}\mathcal{T}$ symmetries, while if $\mathbf{L}$-vector is along any high-symmetry in-plane axis (highlighted in Figure \ref{fig:1}(a)), only $\mathbf{h}$ in-plane is allowed due to the $C_{2y}\mathcal{T}$ and $C_{2x}$ symmetries. This illustrates the influence of the $\mathbf{L}$ orientation on the $\sigma_{AHE}$ through symmetry, and thus the fact that the symmetries of the planes corresponding to $\alpha=$ 0, 60, and 120 deg and $\alpha=$ 30, 90, and 150 deg are different (Figure \ref{fig:5}) can give a plausible explanation to the anisotropy of $\sigma_{AHE}$ in these planes.

\section{Conclusion}
In conclusion, the implications of our measurements on the external field orientation dependence of $\sigma_{AHE}$ in epitaxial Mn$_5$Si$_3$ films are two-fold. First, the systematic changes in $\sigma_{AHE}$ supported by the systematic changes in AMR$_{||}$ upon the field rotation underlines our ability to manipulate the $\mathbf{L}$-orientation with an external field of the order of 2 T and above. This is an important premise for further experimentally feasible spin transport studies in altermagnets where control over the $\mathbf{L}$-orientation is critical. Second, our key result is that the $\sigma_{AHE}$ in epitaxial Mn$_5$Si$_3$ films shows anisotropic behavior as a function of the external field-induced orientation of $\mathbf{L}$ relative to the underlying crystal structure, and this anisotropic behavior depends on the crystal quality and is unlikely to relate to the magnetocrystalline anisotropy. This observation of the anisotropic $\sigma_{AHE}$ adds to the growing array of properties of epitaxial Mn$_5$Si$_3$: theoretical prediction of the collinear compensated hexagonal phase required and experimental demonstrations of vanishing net magnetization, strong spontaneous anomalous Hall \cite{Reichlova2020, Han2024} and Nernst {\cite{Badura2024, Han2024nernst}} effects , and their dependence on the crystallinity- \cite{Kounta2023} and Mn-content\cite{Han2024nernst}. These properties together advocate the altermagnetic phase in this material.

\begin{acknowledgments}
This work was supported by the French national research agency (ANR) and the Deutsche Forschungsgemeinschaft (DFG) (Project MATHEEIAS- Grant No. ANR-20-CE92-0049-01 / DFG-445976410). H. R., E. S., S. T. B. G. and A. T. are supported by the DFG-GACR grant no. 490730630. J. R. acknowledges MINECO for the Margaritas Salas program. D. K. acknowledges the academy of sciences of the Czech Republic for the Lumina quaeruntur program. AB,ES,DK, LS,TJ acknowledge the Ministry of Education, Youth and Sports of the Czech Republic through the OP JAK call Excellent Research (TERAFIT Project No. CZ.02.01.01/00/22\_008/0004594)
\end{acknowledgments}

\newpage

\bibliography{refs.bib}


\end{document}